\documentclass[aps,showpacs,preprintnumbers,amsmath,amssymb]{revtex4}

\usepackage{float}
\usepackage{graphics,epsfig}
\usepackage{graphicx}
\usepackage{dcolumn}
\usepackage{bm}

\begin{document}
\baselineskip=0.8 cm
\title{{\bf Geodetic precession and strong gravitational lensing in the dynamical Chern-Simons modified gravity}}

\author{Songbai Chen}
\email{csb3752@163.com}
\affiliation{Institute of Physics and Department of Physics, Hunan Normal University,  Changsha, Hunan 410081, P. R. China. \\
Key Laboratory of Low Dimensional Quantum Structures and Quantum
Control (Hunan Normal University), Ministry of Education, P. R.
China.}

\author{Jiliang Jing}
\email{jijing@hunnu.edu.cn}
\affiliation{Institute of Physics and Department of Physics, Hunan Normal University,  Changsha, Hunan 410081, P. R. China. \\
Key Laboratory of Low Dimensional Quantum Structures and Quantum
Control (Hunan Normal University), Ministry of Education, P. R.
China.}

\vspace*{0.2cm}
\begin{abstract}
\baselineskip=0.6 cm
\begin{center}
{\bf Abstract}
\end{center}

We have investigated the geodetic precession and the strong
gravitational lensing in the slowly-rotating black hole in the
dynamical Chern-Simons modified gravity theory. We present the
formulas of the orbital period $T$ and the geodetic precession angle
$\Delta\Theta$ for the timelike particles in the circular orbits
around the black hole, which shows that the change of the geodetic
precession angle with the Chern-Simons coupling parameter $\xi$ is
converse to the change of the orbital period with $\xi$ for fixed
$a$. We also discuss the effects of the Chern-Simons coupling
parameter on the strong gravitational lensing when the light rays
pass close to the black hole and obtain that for the stronger
Chern-Simons coupling the prograde photons may be captured more
easily, and conversely, the  retrograde photons is harder to be
captured in the slowly-rotating black hole in the dynamical
Chern-Simons modified gravity. Supposing that the gravitational
field of the supermassive central object of the Galaxy can be
described by this metric, we estimated the numerical values of the
main observables for gravitational lensing in the strong field
limit.
\end{abstract}

\pacs{ 04.70.Dy, 95.30.Sf, 97.60.Lf } \maketitle
\newpage
\section{Introduction}

Since Einstein's general relativity was set forth in the last
century, it has been of interest in the study of possible
modifications of his theory. One of the most promising extension of
general relativity is Chern-Simons modified gravity
\cite{Lue,Jackiw,Alexander}, in which the Einstein-Hilbert action is
modified by adding a parity-violating Chern-Simons term, which
couples to gravity via a scalar field. The parity-violating
Chern-Simons term is a combination of the second order in the
curvature tensor and a Chern-Simons scalar field. Thus, the
Chern-Simons modified gravity is a high-energy extension of
Einstein's general relativity. Actually, Chern-Simons correction is
necessary in the string theory as an anomaly-canceling term to
conserve unitarity \cite{Alexander1,Svrcek,Alvarez,Campbell}. In
loop quantum gravity, it is required to ensure gauge invariance of
the Ashtekar variables \cite{Ashtekar}. Moreover, the Chern-Simons
modified gravity could help us to explain several problems in
cosmology, such as, dark energy and dark matter \cite{Konno2},
baryon asymmetry \cite{Alexander1,Alexander2,Garcia}, and so on.

It is well know that there exist two formulations of Chern-Simons
modified gravity. In the non-dynamical formulation, the Chern-Simons
scalar field is an \textit{a priori} prescribed function so that its
effective evolution equation can be reduced to a differential
constraint on the space of allowed solutions
\cite{Alexander3,Alexander4,Alexander5}. While in the dynamical
formulation, the Chern-Simons scalar field is treated as a dynamical
field possessing its own stress-energy tensor and an evolution
equation \cite{Smith,Yunes1,Konno3}. It must be pointed that
although the non-dynamical Chern-Simons gravity action can be
obtained as a certain limit of the dynamical Chern-Simons gravity
action, the non-dynamical Chern-Simons gravity and dynamical
Chern-Simons gravity are inequivalent and independent theories. In
general, the solutions of the non-dynamical Chern-Simons gravity
cannot be obtained from the solutions of the dynamical Chern-Simons
gravity \cite{Yunes1}.

The characteristic observational signature of the Chern-Simons
modified gravity could allow us to discriminate an effect of this
theory from other phenomena. Since the Schwarzschild solution is
unaffected by Chern-Simons modified gravity, the solar system tests
of general relativity do not put strict bounds on the magnitude of
this correction. Recently, Cardoso \textit{et al.} \cite{Cardoso}
studied the evolution of the dynamical Chern-Simons perturbations in
the background of a Schwarzschild black hole and found that the
quasinormal modes could offer a way to detect the correction from
dynamical Chern-Simons terms.  For a rotating black hole, it is
allowed to possess a non-vanishing Chern-Simons scalar field, which
brings convenient for us to probe the observational signature of the
Chern-Simons modified gravity. Thus, a lot of attention have been
focused on the studies of rotating black holes in the Chern-Simons
modified gravity \cite{Konno2,Alexander4,Alexander5,Yunes1,Konno3,
Ciufolini1,Ciufolini2,Yunes3,Harko,Amarilla,Ahmedov1,Carlos}. In the
non-dynamical formulation, Alexander and Yunes
\cite{Alexander4,Alexander5} adopted to a far-field approximation
and obtained a rotating black hole solution in the Chern-Simons
modified gravity. In the non-dynamical framework, it was found that
the Chern-Simons modified theory predicts an anomalous precession
effect \cite{Alexander5}, which was tested  with LAGEOS
\cite{Smith,Ciufolini1,Ciufolini2}. Using double binary pulsar data,
Yunes and Spergel \cite{Yunes3} obtained a bound on the
non-dynamical model with a canonical Chern-Simons scalar field. In
the dynamical Chern-Simons modified gravity, Yunes and Pretorius
\cite{Yunes1} found a rotating black hole solution by using of the
small-coupling and slow-rotation limit. This rotating black hole
solution was also obtained by Konno \textit{et al.} in
\cite{Konno3}. Harko \cite{Harko} studied the properties of the thin
accretion disk around this black hole and probed the effect of the
Chern-Simons coupling parameter on the flux and the emission
spectrum of the accretion disks. Amarilla \textit{et al.}
\cite{Amarilla} studied the null geodesics corresponding to a
slowly-rotating black hole in the dynamical Chern-Simons gravity and
discussed the effect of the Chern-Simons term on the shadows casted
by a black hole. These results are very useful for us to understand
the properties of black holes in the Chern-Simons modified gravity.
The main purpose of this paper is to study geodetic precession and
the strong gravitational lensing in the slowly-rotating black hole
in the dynamical Chern-Simons modified gravity theory and to see
what effects of the Chern-Simons coupling parameter on the geodetic
precession angle for the timelike particles and the coefficients of
gravitational lensing in the strong field limit.

The plan of our paper is organized as follows. In Sec.II we
introduce briefly the dynamical formulation of the Chern-Simons
modified gravity and present the a slowly-rotating black hole
solution found in \cite{Yunes1,Konno3}. In Sec.III, we calculate the
effects of the Chern-Simons term on the geodetic precession in the
circular orbits around this black hole. In Sec.IV, we adopt to
Bozza's method \cite{Bozza2,Bozza3} and obtain the deflection angles
for light rays passing close to the slowly-rotating black hole in
the dynamical Chern-Simons modified gravity, and then probe the
effects of the Chern-Simons coupling parameter on the deflection
angle and on the coefficients in the strong field limit.  At last,
we present a summary.

\section{Slowly-rotating black hole in the dynamical Chern-Simons modified gravity}

In this section, we introduce briefly the slowly-rotating black hole
in the dynamical Chern-Simons modified gravity \cite{Yunes1,Konno3}.
The action in the dynamical Chern-Simons modified gravity theory can
be expressed as
\begin{eqnarray}
S=\int d^4x\sqrt{-g}\bigg[\kappa R+\frac{\alpha}{4}\vartheta\;
^{*}RR -\frac{\beta}{2}\bigg(g^{\mu\nu}\nabla_{\mu}\vartheta
\nabla_{\nu}\vartheta+V(\vartheta)\bigg)+\mathcal{L}_{matt}\bigg].
\label{action}
\end{eqnarray}
The first term in the right side of Eq.(\ref{action}) is the
standard Einstein-Hilbert term with $\kappa^{-1}=16\pi G$. The
second term denotes the Chern-Simons correction, which consists of
the product of a Chern-Simons scalar field $\vartheta$ and the
Pontryagin density $^{*}RR$, defined via $^{*}RR=
\;^{*}R^{a\;cd}_{\;\;b}R^{b}_{\;\;acd}$. The dual Riemann-tensor $
\;^{*}R^{a\;cd}_{\;\;b}$ is given by
\begin{eqnarray}
^{*}R^{a\;cd}_{\;\;b}=\frac{1}{2}\epsilon^{cdef}R^{a}_{\;\;bef},
\end{eqnarray}
with $\epsilon^{cdef}$ is the 4-dimensional Levi-Civita tensor. The
parameters $\alpha$ and $\beta$ are dimensional coupling constants.
The coupling constant $\beta$ is allowed to be arbitrary in the
dynamical formulation of the Chern-Simons  modified gravity, but it
is set to zero in the non-dynamical framework \cite{Yunes1}.

Varying the action S with respect to the metric and the Chern-Simons
coupling field, one can find that the modified gravitational field
equation and the motion equation of the scalar field $\vartheta$
obey
\begin{eqnarray}
G_{\mu\nu}+\frac{\alpha}{\kappa}C_{\mu\nu}=\frac{1}{2\kappa}(T^{matt}_{\mu\nu}+T^{\vartheta}_{\mu\nu}),\label{grav1}
\end{eqnarray}
and
\begin{eqnarray}
\beta\nabla_{\mu}\nabla^{\mu}\vartheta=\beta\frac{dV(\vartheta)}{d\vartheta}-\frac{\alpha}{4}\;^{*}RR,\label{moto1}
\end{eqnarray}
respectively. Here $G_{\mu\nu}$ is the Einstein tensor and
$C_{\mu\nu}$ is the Cotton tensor. Obviously, the evolution of the
scalar field $\vartheta$ depends not only on its stress-energy
tensor, but also on the curvature of spacetime.

Employing the small-coupling and slow-rotating approximation, one
can  obtain a black-hole solution with non-zero coupling constants
in the dynamical Chern-Simons modified gravity, which can be
expressed as \cite{Yunes1,Konno3}
\begin{eqnarray}
ds^2&=&-g_{tt}dt^2+g_{rr}dr^2+g_{\theta\theta}d\theta^2+g_{\phi\phi}
d\phi^2-2g_{t\phi}dtd\phi, \label{metric0} \\
\vartheta &=&\frac{5}{8}\frac{\alpha}{\beta}\frac{a}{M}\frac{
\cos{\theta}}{r^2}\bigg(1+\frac{2M}{r}+\frac{18M^2}{5r^2}\bigg),
\end{eqnarray}
with
\begin{eqnarray}
g_{tt}&=&1-\frac{2M}{r}+\frac{2a^2M}{r^3}\cos^2{\theta},\nonumber\\
g_{rr}&=&\bigg(1-\frac{2M}{r}\bigg)^{-1}\bigg[1+\frac{a^2}{r}\bigg(\cos^2{\theta}-\bigg(1-\frac{2M}{r}\bigg)^{-1}\bigg)\bigg],\nonumber\\
g_{\theta\theta}&=&r^2+a^2\cos^2{\theta},\nonumber\\
g_{t\phi}&=&\frac{2Ma}{r}\sin^2{\theta}-\frac{5\xi
a}{8r^4}\bigg(1+\frac{12M}{7r}+\frac{27M^2}{10r^2}\bigg)\sin^2{\theta},\nonumber\\
g_{\phi\phi}&=&r^2\sin^2{\theta}+a^2\sin^2{\theta}\bigg(1+\frac{2M}{r}\sin^2{\theta}\bigg).
\end{eqnarray}
Here the parameter $\xi$ is related to the coupling constants
$\alpha$ and $\beta$ by $\xi=\alpha ^2/(\beta\kappa)$, which has an
exact dimension $[L]^4$. As in ref.\cite{Yunes1}, we can define a
dimensionless parameter $\zeta$ by re-scaling by a factor $(2M)^4$,
i.e., $\zeta=\xi/(2M)^4$. If $\alpha$ tends to zero, one can find
that Chern-Simons scalar field $\vartheta$ disappears and the metric
(\ref{metric0}) return to that of the slow-rotating Kerr black hole
in the general relativity. Since the Chern-Simons scalar field
$\vartheta$ has positive energy \cite{Yunes1}, it is natural that
the parameter $\xi$ is non-negative. Here we limit ourselves to the
case where $\xi\geq 0$ and study the effect of $\xi$ on the geodetic
precession and the strong gravitational lensing in the background
(\ref{metric0}).

\section{Geodetic precession in the slowly-rotating black hole in the dynamical Chern-Simons modified
gravity}

The timelike geodesics in the slowly-rotating black hole in
dynamical Chern-Simons modified gravity were considered in
\cite{Harko,Carlos}.  Harko \cite{Harko} studied the properties of
the thin accretion disk around the black hole (\ref{metric0}).
Sopuerta \textit{et al}\cite{Carlos} considered the timelike
geodesic equations for the massive particles and found that the
location of the innermost stable circular orbit (ISCO) and the three
physical fundamental frequencies associated with the time $\tau$ for
the particles are modified in the Chern-Simons modified gravity.
However, in Ref.\cite{Carlos} the geodesic precession of orbits
around Chern-Simons black holes is only illustrated numerically for
a few examples, while an analytic expression for this physical
quantity is still missing. In this paper, we will present an clear
expression of $T$ and study the effects of the Chern-Simons term on
the Kepler's third law,  and then study the geodetic precession of
the massive particles around the black hole (\ref{metric0}).

Let us start with the condition $\theta=\pi/2$, which set the orbits
on the equatorial plane. In this case, one can find that the
timelike geodesics take the form
\begin{eqnarray}
&&u^{t}=\frac{dt}{d\tau}=\frac{Eg_{\phi\phi}-Lg_{t\phi}}{g^2_{t\phi}+g_{tt}g_{\phi\phi}},\label{u1}\\
&&u^{\phi}=\frac{d\phi}{d\tau}=\frac{Eg_{t\phi}+Lg_{tt}}{g^2_{t\phi}+g_{tt}g_{\phi\phi}},\label{u2}\\
&&\bigg(\frac{dr}{d\tau}\bigg)^2+V_{eff}(r)=E^2,
\end{eqnarray}
with the effective potential
\begin{eqnarray}
V_{eff}(r)=\frac{1}{g_{rr}}\bigg(1+\frac{E^2[g_{rr}(g^2_{t\phi}+g_{tt}g_{\phi\phi})-g_{\phi\phi}]+2ELg_{t\phi}+L^2g_{tt}}{g^2_{t\phi}+g_{tt}g_{\phi\phi}}\bigg),
\end{eqnarray}
where $E$ and $L$ are the specific energy and the specific angular
momentum of particles moving in the orbits, respectively. For the
stable circular orbit in the equatorial plane, the effective
potential $V(r)$ must obey
\begin{eqnarray}
V_{eff}(r)=E^2, \;\;\;\;\;\;\;\frac{dV_{eff}(r)}{dr}=0.
\end{eqnarray}
Solving above equations, one can obtain
\begin{eqnarray}
&&E=\frac{g_{tt}+g_{t\phi}\Omega}{\sqrt{g_{tt}+2g_{t\phi}\Omega-g_{\phi\phi}\Omega^2}},\nonumber\\
&&L=\frac{-g_{t\phi}+g_{\phi\phi}\Omega}{\sqrt{g_{tt}+2g_{t\phi}\Omega-g_{\phi\phi}\Omega^2}},\nonumber\\
&&\Omega=\frac{d\phi}{dt}=\frac{g_{t\phi,r}+\sqrt{(g_{t\phi,r})^2+g_{tt,r}g_{\phi\phi,r}}}{g_{\phi\phi,r}},\label{jsd}
\end{eqnarray}
where $\Omega$ is the angular velocity of particle moving in the
orbits. From Eq.(\ref{jsd}), one can obtain Kepler's third law in
the slowly-rotating black-hole spacetime in the dynamical
Chern-Simons modified gravity
\begin{eqnarray}
T^2&=&\frac{4\pi^2}{M}R^3\bigg[1+a\frac{112MR^5-\xi(567M^2+300MR+140R^2)}{56M^{1/2}R^{13/2}}\nonumber\\&+&
\frac{a^2M}{R^3}\bigg(1-\xi\frac{567M^2+300MR
+140R^2}{28MR^5}+\xi^2\frac{(567M^2+300MR +140R^2)^2}{6272
M^2R^{10}}\bigg)+\mathcal{O}(a^3)\bigg],\label{Time}
\end{eqnarray}
where $T$ is the orbital period and $R$ is the radius of the
circular orbit. The later terms in the right hand side is the
correction by the $a$ and the Chern-Simons term. Obviously, the
correction term disappears as $a$ approaches zero. It is reasonable
because that as $a$ vanishes the metric (\ref{metric0}) reduces to
that of the Schwarzschild black hole in the general relativity.
Since the black hole is slowly rotating, the correction is dominated
by the first-order terms in $a$. Thus, when the black hole rotates
in the same direction as the particle, i.e., $a>0$, the orbital
period $T$ decreases with the Chern-Simons coupling parameter $\xi$.
But when the black hole rotates in the converse direction as the
particle, i.e., $a<0$, the orbital period $T$ increases with the
Chern-Simons coupling parameter $\xi$.

Now, we are in the position to study the geodetic precession of a
timelike particle in the circular orbits around the slowly-rotating
black hole in the dynamical Chern-Simons modified gravity. As in
\cite{Ksq}, we regard the rotating axis of the gyroscope carried by
a satellite as a spacelike spin-vector $S^{\mu}$, which parallely
transported along a timelike geodesic with the four-velocity
$u^{\mu}$. Thus, the parallel transporting equation of $S^{\mu}$ in
the direction of $u^{\mu}$ can be expressed as
\begin{eqnarray}
u^{\mu}\nabla_{\mu}S^{\nu}=0.\label{S11}
\end{eqnarray}
From the orthogonality and normalization conditions, one can find
\begin{eqnarray}
u^{\mu}S_{\mu}=0,\;\;\;\;\;\;\;S^{\mu}S_{\mu}=1.\label{con}
\end{eqnarray}
In the slowly-rotating black hole in the dynamical Chern-Simons
modified gravity, the parallel transporting equation (\ref{S11})
along the circular orbits in the equatorial plane reads
\begin{eqnarray}
\baselineskip=1 cm
&&\frac{dS^{t}}{d\tau}+\mathcal{A}S^{r}=0,\label{sge1}\\
&&\frac{dS^{r}}{d\tau}+\mathcal{B}S^{t}+\mathcal{C}S^{\phi}=0,\label{sge2}\\
&&\frac{dS^{\theta}}{d\tau}=0,\label{sge3}\\
&&\frac{dS^{\phi}}{d\tau}+\mathcal{D}S^{r}=0,\label{sge4}
\end{eqnarray}
with
\begin{eqnarray}
\mathcal{A}&&=\frac{1}{2}\bigg[\bigg(\frac{g_{tt,r}g_{\phi\phi}+g_{t\phi,r}g_{t\phi}}{g^2_{t\phi}+g_{tt}g_{\phi\phi}}\bigg)u^t
+\bigg(\frac{g_{t\phi,r}g_{\phi\phi}-g_{t\phi}g_{\phi\phi,r}}{g^2_{t\phi}+g_{tt}g_{\phi\phi}}\bigg)u^{\phi}\bigg]\bigg|_{r=R},\\
\mathcal{B}&&=\frac{1}{2}\bigg[\bigg(\frac{g_{tt,r}}{g_{rr}}\bigg)u^t+\bigg(\frac{g_{t\phi,r}}{g_{rr}}\bigg)u^{\phi}\bigg]\bigg|_{r=R},\\
\mathcal{C}&&=\frac{1}{2}\bigg[\bigg(\frac{g_{t\phi,r}}{g_{rr}}\bigg)u^t-\bigg(\frac{g_{\phi\phi,r}}{g_{rr}}\bigg)u^{\phi}\bigg]\bigg|_{r=R},\\
\mathcal{D}&&=\frac{1}{2}\bigg[\bigg(\frac{g_{tt,r}g_{t\phi}-g_{tt}g_{t\phi,r}}{g^2_{t\phi}+g_{tt}g_{\phi\phi}}\bigg)u^t
+\bigg(\frac{g_{t\phi,r}g_{t\phi}+g_{tt}g_{\phi\phi,r}}{g^2_{t\phi}+g_{tt}g_{\phi\phi}}\bigg)u^{\phi}\bigg]\bigg|_{r=R}.
\end{eqnarray}
Combining above equations with Eqs. (\ref{u1}) and (\ref{u2}), we
can obtain the spin-vector $S^{\mu}$
\begin{eqnarray}
\baselineskip=1 cm
&&S^{t}=C^{t}\sin{(\varpi \tau)},\\
&&S^{r}=C^{r}\cos{(\varpi \tau)},\\
&&S^{\theta}=C^{\theta},\\
&&S^{\phi}=C^{\phi}\sin{(\varpi \tau)},
\end{eqnarray}
with
\begin{eqnarray}
\varpi=\sqrt{-(\mathcal{A}\mathcal{B}+\mathcal{C}\mathcal{D})}.\label{om1}
\end{eqnarray}
Here we have imposed initial condition $S^t=S^{\phi}=0$ at $\tau=0$.
The coefficients $C^{t}$, $C^{r}$, $C^{\theta}$ and $C^{\phi}$ can
be constrained by the orthogonality and normalization conditions
(\ref{con}).

As $\phi$ goes from $0$ to $2\pi$, one can obtain that the proper
time $\tau$ goes from $0$ to $\tau_p= 2\pi/u^{\phi}$. Thus, the
geodetic precession angle $\Delta\Theta$ during one orbital period
can be expressed as
\begin{eqnarray}
\Delta\Theta=|\varpi\tau_p-2\pi|=\bigg|2\pi\bigg(\frac{\varpi}{u^{\phi}}-1\bigg)\bigg|.\label{sc1}
\end{eqnarray}
Substituting (\ref{u2}) and (\ref{om1}) into (\ref{sc1}), one can
expand Eq. (\ref{sc1}) as a power series in $M/R$ up to
$\mathcal{O}(\frac{M^5}{R^5})$ and obtain the geodetic precession
angle $\Delta\Theta$ in the slowly-rotating black hole in dynamical
Chern-Simons modified gravity
\begin{eqnarray}
\Delta\Theta&=&\frac{3\pi
M}{R}\bigg[\bigg(1+\frac{3M}{4R}+\frac{9M^2}{8R^2}+\frac{135M^3}{64R^3}\bigg) \nonumber\\
&&-\frac{a}{\sqrt{MR}}\bigg(\frac{2}{3}+\frac{M}{R}
+\frac{9M^2}{4R^2}+\frac{45M^3}{8R^3}-\frac{5\xi}{6MR^3}\bigg)+\frac{a^2}{R^2}\bigg(\frac{1}{3}+\frac{3M}{2R}+\frac{45M^2}{8R^2}\bigg)+\mathcal{O}(a^3)\bigg].
\label{sc2}
\end{eqnarray}
From Eq. (\ref{sc2}), it is easy to find that the geodetic
precession angle $\Delta\Theta$ increases with the Chern-Simons
coupling parameter $\xi$ if $a>0$ and it decreases with $\xi$ if
$a<0$. Comparing with Eq. (\ref{Time}), we find that the dependent
of the geodetic precession angle on the $\xi$ is converse to the
dependent of the orbital period on the $\xi$. It is understandable
by a fact that the increase of  the orbital period $T$ leads to the
decrease of the angular velocity $\omega$ of particle and then it
results in the decrease of the precession angle $\Delta\Theta$.

\section{Deflection angle in the slowly-rotating black hole in the dynamical Chern-Simons modified
gravity}

The null geodesics was considered to study the properties of the
shadows casted by a slowly-rotating black hole in the dynamical
Chern-Simons gravity \cite{Amarilla}. In this section, we will study
deflection angles of the light rays when they pass close to the
slowly-rotating black hole in dynamical Chern-Simons modified
gravity, and then probe the effects of the Chern-Simons coupling
parameter $\xi$ on the deflection angle and the coefficients in the
strong field limit.

\subsection{Formulas in the strong gravitational lensing}

As in the former, we also consider only the case the light ray is
limited in the equatorial plane.  With this condition,  the reduced
metric for the slowly-rotating black hole in dynamical Chern-Simons
modified gravity can be expressed as
\begin{eqnarray}
ds^2&=&-A(x)dt^2+B(x)dx^2+C(x) d\phi^2-2D(x)dtd\phi, \label{metric1}
\end{eqnarray}
where we adopt to a new radial coordinate $x= r/2M$ and the metric
coefficients  have the form
\begin{eqnarray}
A(x)&=&1-\frac{1}{x},\\
B(x)&=&\frac{x^2-x-\hat{a}^2}{(x-1)^2},\\
C(x)&=&x^2+\frac{\hat{a}^2(x+1)}{x},\\
D(x)&=&\frac{\hat{a}}{x}-\frac{\hat{a}\zeta(280x^2+189x+240)}{448x^6}.
\end{eqnarray}
Here the quantities $\zeta=\frac{\xi}{(2M)^4}$ and
$\hat{a}=\frac{a}{2M}$ are the re-scaled Chern-Simons coupling
parameter and the re-scaled rotation parameter of black hole,
respectively. Obviously, the parameters $\zeta$ and $\hat{a}$ are
dimensionless. For simplicity, we set $2M=1$ in the following
calculations.

As in Ref. \cite{Bozza3}, the null geodesics take the form
\begin{eqnarray}
&&\frac{dt}{d\lambda}=\frac{C(x)-JD(x)}{D(x)^2+A(x)C(x)},\label{u3}\\
&&\frac{d\phi}{d\lambda}=\frac{D(x)+JA(x)}{D(x)^2+A(x)C(x)},\label{u4}
\end{eqnarray}
where $\lambda$ is an affine parameter along the geodesics and $J$
is the angular momentum of the photon.

For the null geodesics, the Lagrangian
$\mathcal{L}=\frac{1}{2}g_{\mu\nu}\dot{x^{\mu}}\dot{x^{\nu}}$
vanishes. This implies that
\begin{eqnarray}
\dot{x}=\pm\sqrt{\frac{C(x)-J[2D(x)+JA(x)]}{B(x)[D(x)^2+A(x)C(x)]}}.
\end{eqnarray}
Clearly, $\dot{x}$ is equal to zero at the minimum distance of
approach of the light ray. Combining with Eqs. (\ref{u3}) and
(\ref{u4}), one can obtain that \cite{Bozza3}
\begin{eqnarray}
J=u=\frac{-D(x_0)+\sqrt{A(x_0)C(x_0)+D^2(x_0)}}{A(x_0)}.
\end{eqnarray}
where $x_0$ is the closest approach distance and $u$ is the impact
parameter. In the slowly-rotating black-hole spacetime in the
dynamical Chern-Simons modified gravity,  the photon-sphere equation
is given by
\begin{eqnarray}
A(x)C'(x)-A'(x)C(x)+2J[A'(x)D(x)-A(x)D'(x)]=0.\label{root}
\end{eqnarray}
Obviously, this equation is more complex than that in the background
of a static and spherical black hole \cite{Vir2}. It is difficult to
obtain an analytical form for the photon-sphere radius in this case.
However, we can expand Eq.(\ref{root}) as a power series in
$\hat{a}$ and find that the photon-sphere radius in the
slowly-rotating approximation can be expressed as
\begin{eqnarray}
x_{ps}=\frac{3}{2}-\hat{a}\bigg(\frac{2\sqrt{3}}{3}-\frac{62\sqrt{3}}{243}\zeta\bigg)
-\hat{a}^2\bigg(\frac{4}{9}-\frac{17924}{5103}\zeta+\frac{896024}{413343}\zeta^2\bigg)+\mathcal{O}(\hat{a}^3).\label{rps}
\end{eqnarray}
\begin{figure}[ht]
\begin{center}
\includegraphics[width=7cm]{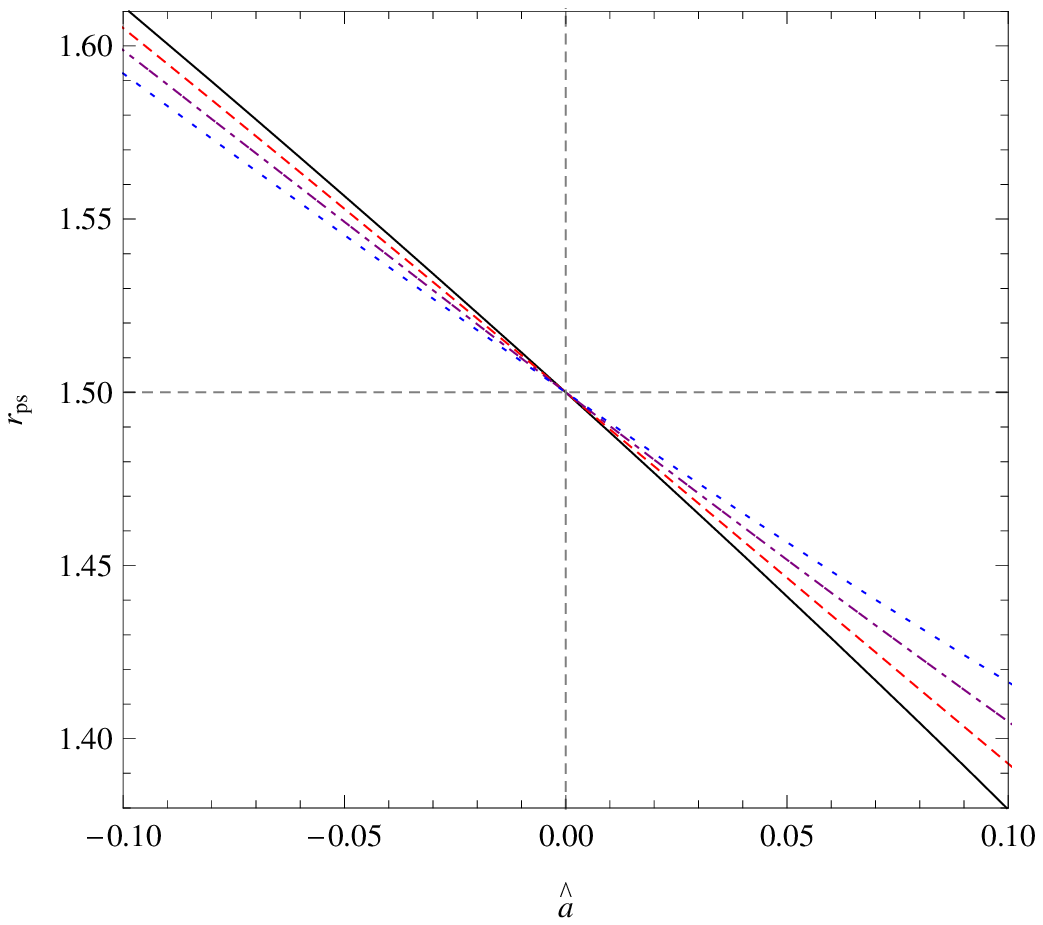}
\caption{Variety of the quantity $r_{ps}=2M x_{ps}$ with the
Chern-Simons coupling parameter $\zeta$ in the slowly-rotating black
hole in the dynamical Chern-Simons modified gravity. The solid,
dashed, dash-dotted and dotted curves are for $\zeta=0$, $0.1$,
$0.2$ and $0.3$, respectively. Here we set $2M=1$.} \label{f1}
\end{center}
\end{figure}
From Eq.(\ref{rps}), one can obtain that the photon-sphere radius
$x_{ps}$ increases with the parameter $\zeta$ if the photons are
winding in the same direction of the black-hole rotation (i.e.,
$\hat{a}>0$), while the radius $x_{ps}$ decreases with the parameter
$\zeta$ if the photons rotate in converse direction to the black
hole (i.e., $\hat{a}<0$). This is also shown in Fig.(\ref{f1}) in
which we plotted the variety of the photon-sphere radius $x_{ps}$
with the Chern-Simons coupling parameter $\zeta$ and the rotating
parameter $\hat{a}$ by solving Eq. (\ref{root}) numerically.
Moreover, we find that as $\zeta\rightarrow 0$, the photon-sphere
radius $x_{ps}$ reduces to that in Kerr black hole. As the rotation
parameter $\hat{a}$ tends to zero, $x_{ps}$ is independent of the
Chern-Simons coupling parameter $\zeta$.

Following ref. \cite{Ein1},  we can obtain the deflection angle for
the photon coming from infinite in the slowly-rotating black hole in
the dynamical Chern-Simons modified gravity.
\begin{eqnarray}
\alpha(x_{0})=I(x_{0})-\pi,
\end{eqnarray}
 and $I(x_0)$ is
\begin{eqnarray}
I(x_0)=2\int^{\infty}_{x_0}\frac{\sqrt{B(x)|A(x_0)|}[D(x)+JA(x)]dx}{\sqrt{D^2(x)+A(x)C(x)}
\sqrt{A(x_0)C(x)-A(x)C(x_0)+2J[A(x)D(x_0)-A(x_0)D(x)]}},\label{int1}
\end{eqnarray}
As in the Schwarzschild black hole spacetime, the deflection angle
increases when parameter $x_0$ decreases. For a certain value of
$x_0$ the deflection angle becomes $2\pi$, so that the light ray
makes a complete loop around the black hole before reaching the
observer. If $x_0$ is equal to the radius of the photon sphere
$x_{ps}$, one can find that the deflection angle diverges and the
photon is captured on a circular orbit. From the above discussion
about the variety of the photon-sphere radius $r_{ps}$ with  the
Chern-Simons parameter $\zeta$, it is easy to obtain that for the
larger $\zeta$ the prograde photons may be captured more easily, and
conversely, the retrograde photons is harder to be captured.

In order to find the behavior of the deflection angle very close to
the photon sphere, we adopt to the evaluation method for the
integral (\ref{int1}) proposed by Bozza \cite{Bozza2}, which has
been widely used in studying of the strong gravitational lensing of
various black holes
\cite{Vir2,Gyulchev,Gyulchev1,Darwin,Vir,Vir1,Vir3,Fritt,Bozza1,Eirc1,whisk,Bhad1,Song1,Song2,
TSa1,AnAv}. Let us now to define a variable
\begin{eqnarray}
z=1-\frac{x_0}{x},
\end{eqnarray}
and rewrite the Eq.(\ref{int1}) as
\begin{eqnarray}
I(x_0)=\int^{1}_{0}R(z,x_0)f(z,x_0)dz,\label{in1}
\end{eqnarray}
with
\begin{eqnarray}
R(z,x_0)&=&2\frac{1-A(x_0)}{A'(z)\sqrt{C(z)}}\frac{\sqrt{B(z)|A(x_0)|}[D(z)+JA(z)]}{\sqrt{D^2(z)+A(z)C(z)}},
\end{eqnarray}
\begin{eqnarray}
f(z,x_0)&=&\frac{1}{\sqrt{A(x_0)-A(z)\frac{C(x_0)}{C(z)}+\frac{2J}{C(z)}[A(z)D(x_0)-A(x_0)D(z)]}}.
\end{eqnarray}
The function $R(z, x_0)$ is regular for all values of $z$ and $x_0$.
However, the function $f(z, x_0)$ diverges as $z$ tends to zero,
i.e., as the photon approaches the photon sphere. Thus, we can split
the integral (\ref{in1}) into the divergent part $I_D(x_0)$ and the
regular one $I_R(x_0)$
\begin{eqnarray}
I_D(x_0)&=&\int^{1}_{0}R(0,x_{ps})f_0(z,x_0)dz, \nonumber\\
I_R(x_0)&=&\int^{1}_{0}[R(z,x_0)f(z,x_0)-R(0,x_{ps})f_0(z,x_0)]dz
\label{intbr}.
\end{eqnarray}
\begin{figure}[ht]
\begin{center}
\includegraphics[width=7cm]{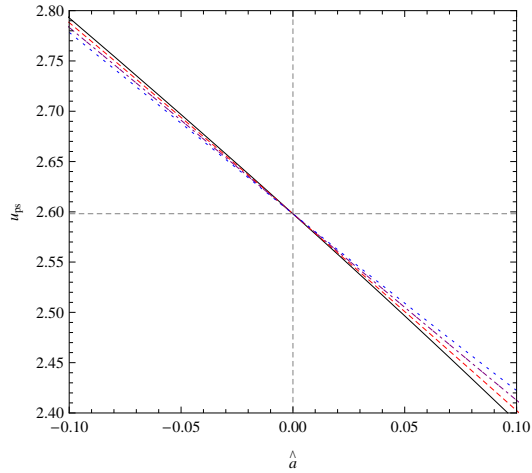}
\caption{The  minimum impact parameter $u_{ps}$ changes with the
Chern-Simons coupling parameter $\zeta$ in the slowly-rotating black
hole in the dynamical Chern-Simons modified gravity. The solid,
dashed, dash-dotted and dotted curves are for $\zeta=0$, $0.1$,
$0.2$ and $0.3$, respectively. Here we set $2M=1$. }\label{fs2}
\end{center}
\end{figure}
Expanding the argument of the square root in $f(z,x_0)$ to the
second order in $z$, we have
\begin{eqnarray}
f_0(z,x_0)=\frac{1}{\sqrt{p(x_0)z+q(x_0)z^2}},
\end{eqnarray}
where
\begin{eqnarray}
p(x_0)&=&\frac{1-A(x_0)}{A'(x_0)C(x_0)}\bigg\{A(x_0)C'(x_0)-A'(x_0)C(x_0)+2J[A'(x_0)D(x_0)-A(x_0)D(x_0)]\bigg\},  \nonumber\\
q(x_0)&=&\frac{(1-A(x_0))^2}{2A'^3(x_0)C^2(x_0)}\bigg(2C(x_0)C'(x_0)A'^2(x_0)+[C(x_0)C''(x_0)-2C'^2(x_0)]A(x_0)A'(x_0)\nonumber\\
&&-C(x_0)C'(x_0)A(x_0)A''(x_0)+
2J\{A(x_0)C(x_0)[A''(x_0)D'(x_0)-A'(x_0)D''(x_0)]\nonumber\\&&+2A'(x_0)C'(x_0)[A(x_0)D'(x_0)-A'(x_0)D(x_0)]\}\bigg).\label{al0}
\end{eqnarray}
Comparing Eq.(\ref{root}) with Eq.(\ref{al0}), one can find that if
$x_{0}$ approaches to the radius of photon sphere $x_{ps}$ the
coefficient $p(x_{0})$ vanishes and the leading term of the
divergence in $f_0(z,x_{0})$ is $z^{-1}$. This means that the
integral (\ref{in1}) diverges logarithmically. The coefficient
$q(x_0)$ takes the form
\begin{eqnarray}
q(x_{ps})&=&\frac{(1-A(x_{ps}))^2}{2A'^2(x_{ps})C(x_{ps})}\bigg\{A(x_{ps})C''(x_{ps})-A''(x_{ps})C(x_{ps})
2J[A''(x_{ps})D(x_{ps})-A(x_{ps})D''(x_{ps})]\bigg\}.
\end{eqnarray}
Therefore the deflection angle in the strong field region can be
expanded in the form \cite{Bozza2}
\begin{eqnarray}
\alpha(\theta)=-\bar{a}\log{\bigg(\frac{u}{u_{ps}}-1\bigg)}+\bar{b}+\mathcal{O}(u-u_{ps}),
\label{alf1}
\end{eqnarray}
with
\begin{eqnarray}
&\bar{a}&=\frac{R(0,x_{ps})}{\sqrt{q(x_{ps})}}, \nonumber\\
&\bar{b}&= -\pi+b_R+\bar{a}\log{\bigg\{\frac{2q(x_{ps})C(x_{ps})}{u_{ps}A(x_{ps})[D(x_{ps}+JA(x_{ps})]}\bigg\}}, \nonumber\\
&b_R&=I_R(x_{ps}), \nonumber\\
&u_{ps}&=\frac{-D(x_{ps})+\sqrt{A(x_{ps})C(x_{ps})+D^2(x_{ps})}}{A(x_{ps})}.\label{coa1}
\end{eqnarray}
\begin{figure}[ht]
\begin{center}
\includegraphics[width=7cm]{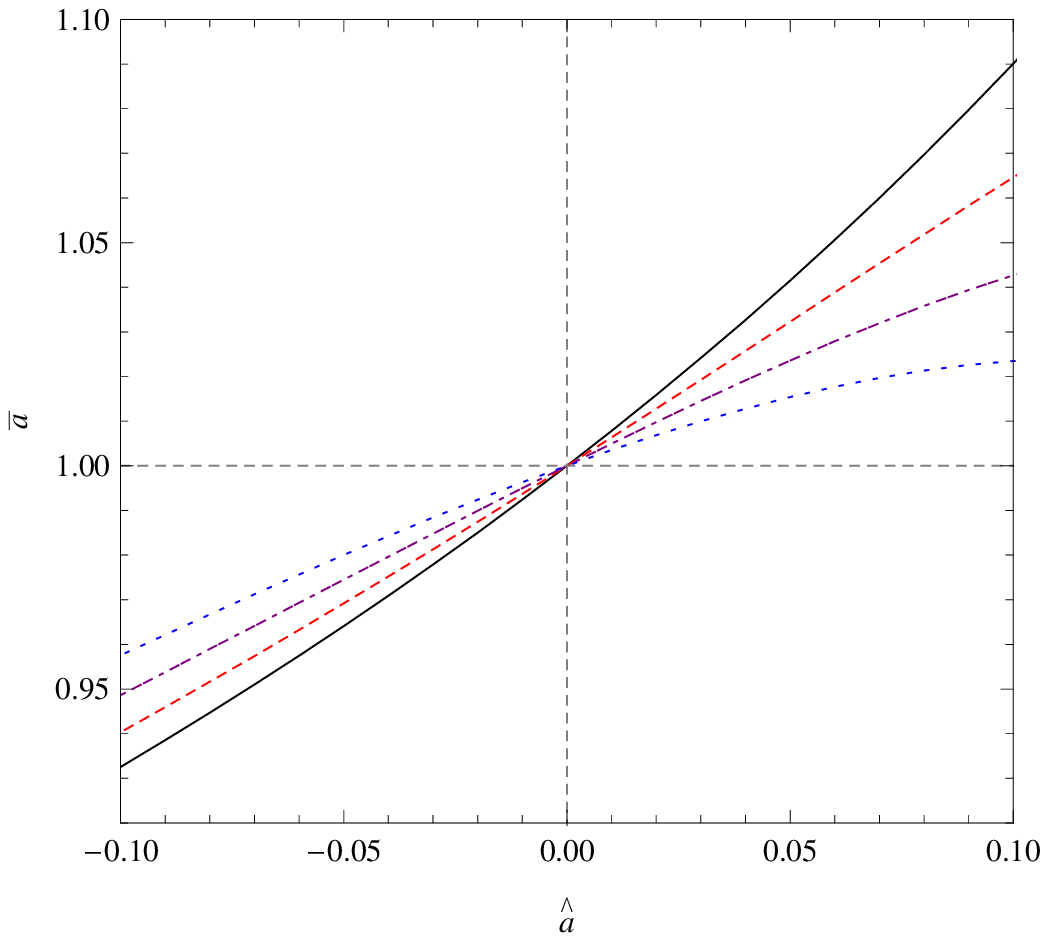}\;\;\;\includegraphics[width=7cm]{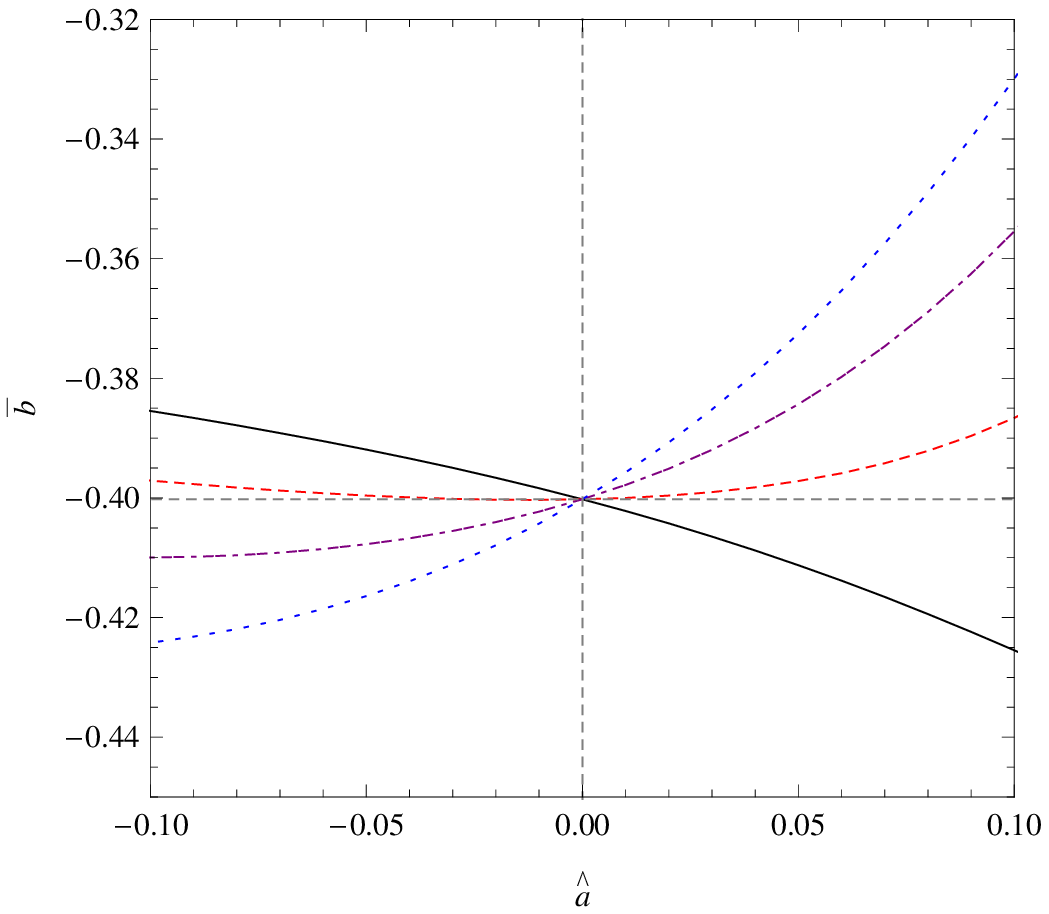}
\caption{Variation of the coefficients of the strong field limit
$\bar{a}$ (the left) and $\bar{b}$ (the right) with the Chern-Simons
coupling parameter $\zeta$ in the slowly-rotating black hole in the
dynamical Chern-Simons modified gravity. The solid, dashed,
dash-dotted and dotted curves are for $\zeta=0$, $0.1$, $0.2$ and
$0.3$, respectively. Here we set $2M=1$. }\label{f3}
\end{center}
\end{figure}
\begin{figure}[ht]
\begin{center}
\includegraphics[width=7cm]{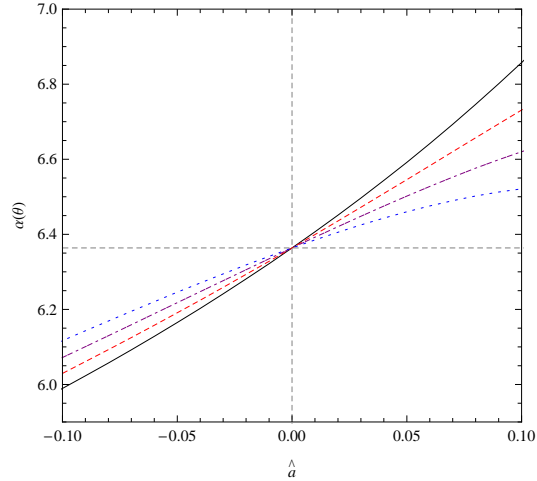}
\caption{Deflection angles evaluated at $u=u_{ps}+0.003$ is a
function of the Chern-Simons coupling parameter $\zeta$ in the
slowly-rotating black hole in the dynamical Chern-Simons modified
gravity. The solid, dashed, dash-dotted and dotted curves are for
$\zeta=0$, $0.1$, $0.2$ and $0.3$, respectively. Here we set
$2M=1$.}\label{f4}
\end{center}
\end{figure}
Making use of Eqs.(\ref{alf1}) and (\ref{coa1}), we can study the
properties of strong gravitational lensing in the slowly-rotating
black hole in the dynamical Chern-Simons modified gravity.
Neglecting terms of order $\mathcal{O}(\hat{a}^3)$ and higher term,
we can expand the the coefficients ($\bar{a}$ and $\bar{b}$ ) in the
strong gravitational lensing and the minimum impact parameter
$u_{ps}$ as a power series in $\hat{a}$
\begin{eqnarray}
&\bar{a}&=1+\frac{2\sqrt{3}\hat{a}(1134-2003\zeta)}{5103}+
\hat{a}^2\bigg(\frac{10}{9}-\frac{133288\zeta}{15309}+\frac{57184906\zeta^2}{8680203}\bigg)+
\mathcal{O}(\hat{a}^3), \nonumber\\
&\bar{b}&=-0.40023-(0.190505-2.06119\zeta)\hat{a}-\hat{a}^2(14.903\zeta^2-13.9387\zeta+0.541507)+\mathcal{O}(\hat{a}^3), \nonumber\\
&b_R&=\log{6}+
\frac{\sqrt{3}\hat{a}}{3}\bigg[\frac{4(1+\log{6})}{3}+\frac{\zeta
(3754-4006\log{6})}{1701}\bigg]\nonumber\\&&+
\hat{a}^2\bigg[\frac{4}{3}+\frac{10}{9}\log{6}+\zeta\bigg(\frac{7528}{1701}-\frac{133288\log{6}}{15309}\bigg)-
\zeta^2\bigg(\frac{67484654}{8680203}-\frac{57184906
\log{6}}{8680203}\bigg)\bigg]
+\mathcal{O}(\hat{a}^3), \nonumber\\
&u_{ps}&
=\frac{3\sqrt{3}}{2}-\hat{a}\bigg(2-\frac{131\zeta}{189}\bigg)-\frac{\sqrt{3}\hat{a}^2}{3}\bigg(1-\frac{2948\zeta}{567
}+\frac{701941\zeta^2}{321489}\bigg)+\mathcal{O}(\hat{a}^3).\label{coa21}
\end{eqnarray}
In figs.(\ref{fs2})-(\ref{f3}), we plotted numerically the changes
of the minimum impact parameter $u_{ps}$ and the coefficients
($\bar{a}$ and $\bar{b}$ ) with $\hat{a}$ and $\zeta$ in the
slowly-rotating black hole in the dynamical Chern-Simons modified
gravity. From Eqs.(\ref{coa21}) and figs. (\ref{fs2})-(\ref{f3}), we
find that the minimum impact parameter $u_{ps}$ and the coefficients
($\bar{a}$ and $\bar{b}$ ) in the strong field limit are functions
of the rotation parameter $\hat{a}$ and the Chern-Simons coupling
parameter $\zeta$. The minimum impact parameter has similar behavior
as the radius of the photon sphere $x_{ps}$. The coefficient
$\bar{a}$ increases with $\hat{a}$ for fixed $\zeta$. For fixed
$\hat{a}$, the coefficient $\bar{a}$ decreases with the Chern-Simons
coupling parameter $\zeta$ if $\hat{a}>0$ and it increases if
$\hat{a}<0$. The coefficient $\bar{b}$ decreases with $\hat{a}$ for
the smaller $\zeta$, but it increases with $\hat{a}$ for the larger
$\zeta$. For fixed $\hat{a}$, the variety of $\bar{b}$ with $\zeta$
is converse to the variety of $\bar{a}$ with $\zeta$. In fig.
(\ref{f4}), we plotted the change of the deflection angles evaluated
at $u=u_{ps}+0.003$ with $\zeta$. It is shown that in the strong
field limit the deflection angles have the similar properties of the
coefficient $\bar{a}$. This means that the deflection angles of the
light rays are dominated by the logarithmic term in the strong
gravitational lensing. Moreover, we also find that for larger
$\zeta$ the deflection angle is larger for the retrograde photon,
while it is smaller for the prograde photon. These imply that for
the larger $\zeta$ the prograde photons may be captured more easily,
and conversely, the  retrograde photons is harder to be captured in
the slowly-rotating black hole in the dynamical Chern-Simons
modified gravity.

\subsection{Observables in the strong deflection limit}

Let us now to study the effect of the Chern-Simons parameter $\zeta$
on the observational gravitational lensing parameters. We start by
assuming that the gravitational field of the supermassive black hole
at the Galactic center of Milky Way can be described by the
slowly-rotating black hole in the dynamical Chern-Simons modified
gravity, and then estimate the numerical values for the main
observables of gravitational lensing in the strong field limit.

In the strong deflection limit, from the lensing geometry we can
rewrite the lens equation as \cite{Bozza3}
\begin{eqnarray}
\gamma=\frac{D_{OL}+D_{LS}}{D_{LS}}\theta-\alpha(\theta) \; mod
\;2\pi
\end{eqnarray}
where $D_{LS}$ is the lens-source distance and $D_{OL}$ is the
observer-lens distance. $\gamma$ is the angle between the direction
of the source and the optical axis. $\theta=u/D_{OL}$ is the angular
separation between the lens and the image.

Following ref.\cite{Bozza3}, one can find that the angular
separation between the lens and the n-th relativistic image is
\begin{eqnarray}
\theta_n\simeq\theta^0_n\bigg(1-\frac{u_{ps}e_n(D_{OL}+D_{LS})}{\bar{a}D_{OL}D_{LS}}\bigg),
\end{eqnarray}
with
\begin{eqnarray}
\theta^0_n=\frac{u_{ps}}{D_{OL}}(1+e_n),\;\;\;\;\;\;e_{n}=e^{\frac{\bar{b}+|\gamma|-2\pi
n}{\bar{a}}}.
\end{eqnarray}
The quantity $\theta^0_n$ is the image positions corresponding to
$\alpha=2n\pi$, and $n$ is an integer. According to the past
oriented light ray which starts from the observer and finishes at
the source the resulting images stand on the eastern side of the
black hole for direct photons ($\hat{a}>0$) and are described by
positive $\gamma$. Retrograde photons ($\hat{a}<0$) have images on
the western side of the black hole and are described by negative
values of $\gamma$.

In the limit $n\rightarrow \infty$, we find that $e_n\rightarrow 0$,
and then the relation between the minimum impact parameter $u_{ps}$
and the asymptotic position of a set of images $\theta_{\infty}$ can
be simplified as
\begin{eqnarray}
u_{ps}=D_{OL}\theta_{\infty}.\label{uhs1}
\end{eqnarray}
In order to obtain the coefficients $\bar{a}$ and $\bar{b}$, one
needs to separate at least the outermost image from all the others.
As in Refs.\cite{Bozza2,Bozza3},  we consider here the simplest case
in which only the outermost image $\theta_1$ is resolved as a single
image and all the remaining ones are packed together at
$\theta_{\infty}$. Thus the angular separation between the first
image and other ones can be expressed as
\cite{Bozza2,Bozza3,Gyulchev1}
\begin{eqnarray}
s=\theta_1-\theta_{\infty}=\theta_{\infty}e^{\frac{\bar{b}-2\pi}{\bar{a}}}.\label{ss1}
\end{eqnarray}
Through measuring $s$ and $\theta_{\infty}$, we can obtain the
strong deflection limit coefficients $\bar{a}$, $\bar{b}$ and the
minimum impact parameter $u_{ps}$. Comparing their values with those
predicted by the theoretical models, we can obtain information about
the parameters of the lens object stored in them.

The mass of the central object of our Galaxy is estimated recently
to be $4.4\times 10^6M_{\odot}$ \cite{Genzel1} and its distance is
around $8.5kpc$, so that the ratio of the mass to the distance
$M/D_{OL} \approx2.4734\times10^{-11}$.  Making use of Eqs.
(\ref{coa21}), (\ref{uhs1}) and  (\ref{ss1})  we can estimate the
values of the coefficients and observables for gravitational lensing
in the strong field limit. For the different $\zeta$ and $\hat{a}$,
the numerical value for the angular position of the relativistic
images $\theta_{\infty}$ and the angular separation $s$ are listed
in the Table I. The dependence of these observables on the
parameters $\zeta$ and $\hat{a}$ are also shown in Fig. (5).
\begin{figure}[ht]
\begin{center}
\includegraphics[width=6cm]{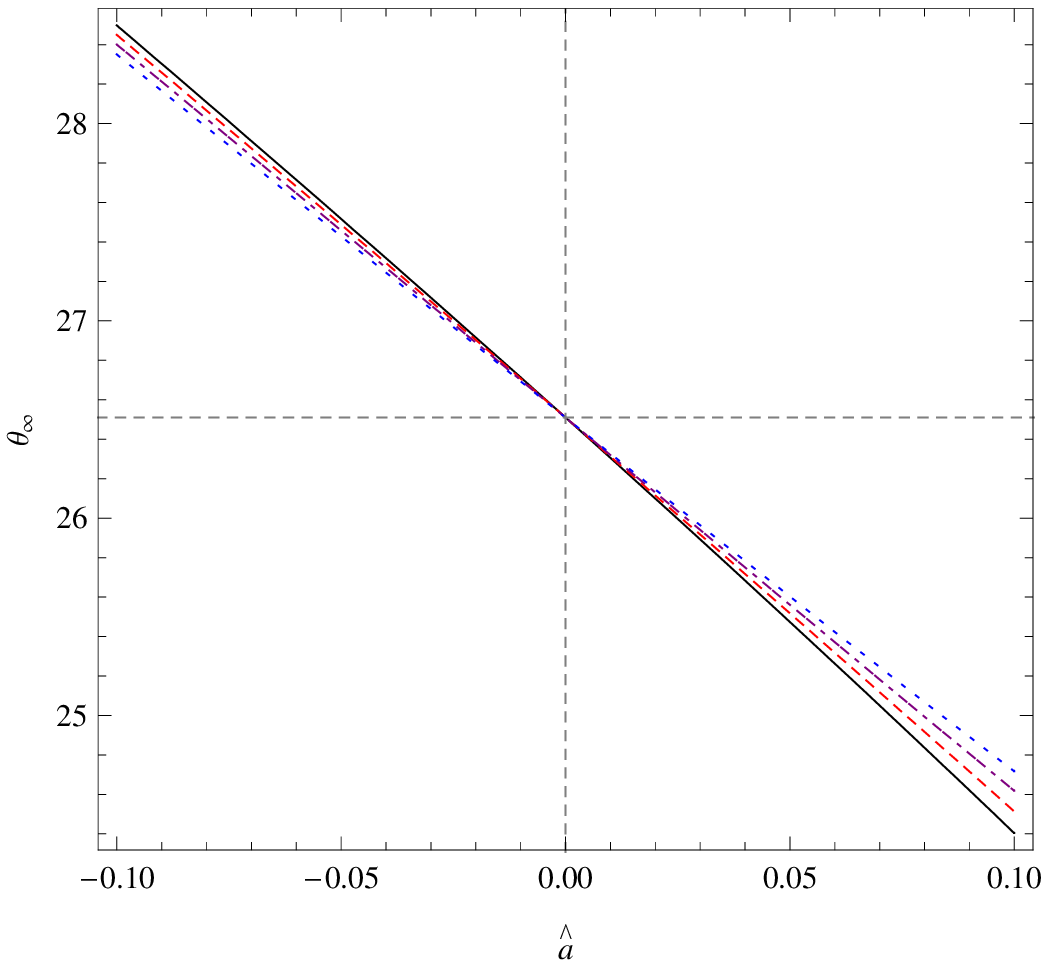}\;\;\;\;\;\includegraphics[width=6cm]{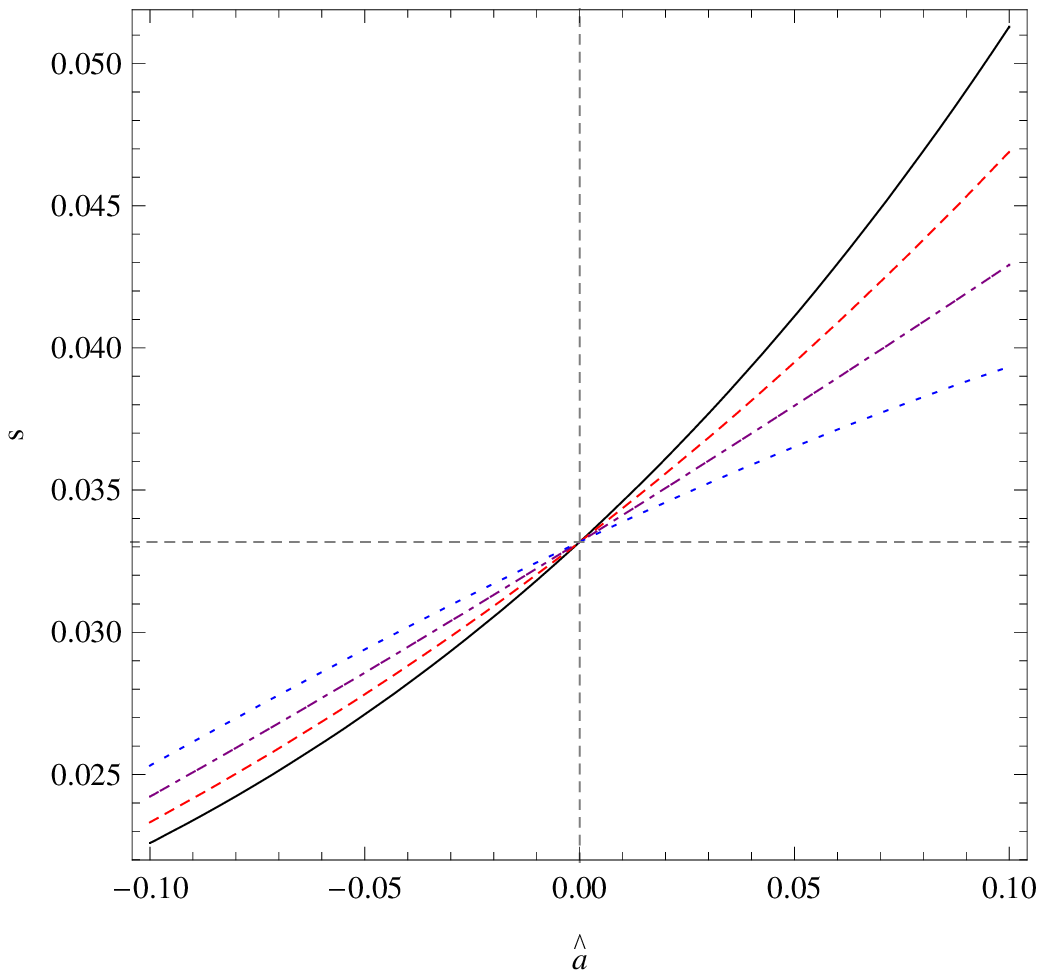}
\caption{Gravitational lensing by the Galactic center black hole.
Variation of the values of the angular position $\theta_{\infty}$,
the angular separation $s$ with parameter $\hat{a}$ in the
slowly-rotating black hole in the dynamical Chern-Simons modified
gravity. The solid, dashed, dash-dotted and dotted curves are for
$\zeta=0$, $0.1$, $0.2$ and $0.3$, respectively.}
\end{center}
\label{51}
\end{figure}
\begin{table}[h]
\begin{center}
\begin{tabular}{|c|c|c|c|c|c|c|c|c|}
\hline \hline &\multicolumn{4}{c|}{$\theta_{\infty}$($\mu$
arcsec)}&\multicolumn{4}{c|}{$s$($\mu$ arcsec)} \\
\hline $\hat{a}$ &$\zeta=0$&$\zeta=0.1$&$\zeta=0.2$ &$\zeta=0.3$&$\zeta=0$&$\zeta=0.1$&$\zeta=0.2$ &$\zeta=0.3$  \\
\hline
 -0.10& 28.497&28.449&28.400&28.350&0.0226& 0.0233 &0.0242&0.0253\\
 \hline
-0.05&27.516&27.487&27.458&27.428&0.0271&0.0278&0.0286&0.0294\\
\hline
0& 26.510&26.510&26.510&26.510&0.0332&0.0332&0.0332&0.0332 \\
\hline
0.05& 25.474&25.518&25.561&25.603&0.0411&0.0395&0.0380&0.0365\\
 \hline
0.10&24.403&24.514&24.618&24.717&0.0513&0.0469&0.0429&0.0393
 \\
\hline\hline
\end{tabular}
\end{center}
\label{tab1} \caption{Numerical estimation for main observables in
the strong field limit for the black hole at the center of our
Galaxy, which is supposed to be described by in the slowly-rotating
black hole in the dynamical Chern-Simons modified gravity.}
\end{table}
Obviously, the observables $\theta_{\infty}$ and $s$ are independent
of the parameter $\zeta$ as the rotation parameter $\hat{a}=0$. From
Table I and Fig. (5), we find that with the increase of $\zeta$, the
angular position of the relativistic images $\theta_{\infty}$
increases for the direct photons $(\hat{a}>0)$ and decrease for the
retrograde photons $(\hat{a}<0)$. The change of the angular
separation $s$ with $\zeta$ is converse to that of
$\theta_{\infty}$.

Theoretically, we could detect the effects of parameter $\zeta$ on
the strong gravitational lensing through the astronomical
observations and then make a constraint on the parameter $\xi$. From
Table I, we find that the values for $\theta_{\infty}$ is very
small, which leads to constrain the parameter $\xi$ more
difficultly. If one can constrain $\zeta<0.1$ with lensing
observations, it is easy to obtain that the bound of the
Chern-Simons coupling parameter is
\begin{eqnarray}
\xi^{1/4}=\zeta^{1/4}2M<8.70\times10^6\text{km},
\end{eqnarray}
which is not stronger than that from the binary pulsar PSR
J0737-3039 A/B \cite{Burgay} obtained by Yunes \textit{et al}
\cite{Yunes1}
\begin{eqnarray}
\xi^{1/4}<1.5\times10^4\text{km}.
\end{eqnarray}
In order that the strong gravitational lensing bound can beat the
binary pulsar one, we would have to constrain $\zeta\sim 8.44\times
10^{-13}$, which seems impossible in the near future.

\section{summary}

In this paper we have extensively studied the geodetic precession
and the strong gravitational lensing in the slowly-rotating black
hole in the dynamical Chern-Simons modified gravity theory. We
present the formulas of the orbital period $T$ and the geodetic
precession angle $\Delta\Theta$ for the timelike particles in the
circular orbits around the black hole. When the black hole rotates
in the same direction as the particle ( $a>0$), the orbital period
$T$ decreases with the Chern-Simons coupling parameter $\xi$. While
the black hole rotates in the converse direction as the particle
($a<0$), the orbital period $T$ increases with the parameter $\xi$.
Moreover, it is shown that the change of the geodetic precession
angle with the $\xi$ is converse to the change of the orbital period
with the $\xi$. We also discuss the effects of the Chern-Simons
coupling parameter on the strong gravitational lensing when the
light rays pass close to the black hole. We find that the
photon-sphere radius, the minimum impact parameter, and the
coefficients in the strong field limit depend on the Chern-Simons
coupling parameter. With the increase of $\zeta$ (i.e., the
re-scaling Chern-Simons coupling parameter $\zeta=\xi/(2M)^4$) the
deflection angle increases for the prograde photon, while it
decreases for the retrograde photon. It means that for the larger
$\zeta$ the prograde photons may be captured more easily, and
conversely, the  retrograde photons is harder to be captured in the
slowly-rotating black hole in dynamical Chern-Simons modified
gravity.

The model was applied to the supermassive black hole in the Galactic
center. Our results show that with the increase of the parameter
$\zeta$ the angular position of the relativistic images
$\theta_{\infty}$ increases for the direct photons $(a>0)$ and
decrease for the retrograde photons $(a<0)$. The change of the
angular separation $s$ with $\zeta$ is converse to that of
$\theta_{\infty}$. Our result also show that the bound of $\xi$ from
the strong gravitational lensing is not stronger than that from the
binary pulsar PSR J0737-3039 A/B \cite{Burgay}. In particular, for
the case $\zeta = 0.1$, we can get the Chern-Simons coupling
parameter $\xi\sim 8.70\times10^6 $ km, which is ruled out by binary
pulsar constraints. If one were to choose the parameter $\xi$ small
enough not to be ruled out by these constraints, the effect would be
much smaller than microarcseconds. Perhaps with the development of
technology, the effects of parameter $\zeta$ on gravitational
lensing may be detected in the future. It would be of interest to
study the Hawking radiation and quasinormal modes of the
slowly-rotating black hole in dynamical Chern-Simons modified
gravity. Work in this direction will be reported in the future.

\begin{acknowledgments}
We thank the referees for their their constructive comments and
suggestions, which make the meaning expressed in our manuscript more
clear. This work was  partially supported by the National Natural
Science Foundation of China under Grant No.10875041,  the Program
for Changjiang Scholars and Innovative Research Team in University
(PCSIRT, No. IRT0964) and the construct program of key disciplines
in Hunan Province. J. Jing's work was partially supported by the
National Natural Science Foundation of China under Grant
Nos.10875040 and No.10935013; 973 Program Grant No. 2010CB833004 and
the Hunan Provincial Natural Science Foundation of China under Grant
No.08JJ3010.

\end{acknowledgments}


\vspace*{0.2cm}
\begin{thebibliography}{99}
\baselineskip=0.6 cm

\bibitem{Lue}Lue A, Wang L M  and Kamionkowski M 1999 Phys. Rev. Lett. {\bf 83}
1506

\bibitem{Jackiw}Jackiw R and Pi S Y 2003 Phys. Rev. D {\bf 68} 104012

\bibitem{Alexander}Alexander S and Yunes N 2009 Phys. Rept. {\bf 480} 1

\bibitem{Alexander1}Alexander  S H S, Gates J and  James S 2006 JCAP {\bf 0606} 018


\bibitem{Svrcek}Svrcek P and Witten E 2006 JHEP {\bf 0606} 051

\bibitem{Alvarez}Alvarez-Gaume L and Witten E 1984 Nucl. Phys. B {\bf 234} 269

\bibitem{Campbell}Campbell  B A, Kaloper N, Madden R and Olive K A 1993 Nucl.
Phys. B {\bf 399} 137


\bibitem{Ashtekar}Ashtekar A, Balachandran A P and Jo S 1989 Int. J. Mod. Phys. A {\bf 4}
1493


\bibitem{Konno2}Konno K, Matsuyama T, Asano Y and Tanda S 2008 Phys. Rev. D
{\bf78} 024037

\bibitem{Alexander2}Alexander S, Peskin M E and Sheikh-Jabbari M M 2006 Phys.
Rev. Lett. {\bf 96} 081301

\bibitem{Garcia} Garcia-Bellido J, Garcia-Perez M and Gonzalez-Arroyo A 2004 Phys. Rev. D {\bf 69} 023504



\bibitem{Alexander3} Alexander S, Finn L S and Yunes N 2008 Phys. Rev. D {\bf 78}  066005


\bibitem{Alexander4} Alexander S and Yunes N 2007 Phys. Rev. Lett. {\bf99}  241101


\bibitem{Alexander5}Alexander S and Yunes N 2007 Phys. Rev. D {\bf 75}  124022





\bibitem{Smith} Smith T L, Erickcek A L, Caldwell R R and
Kamionkowski M 2008 Phys. Rev. D {\bf 77} 024015

\bibitem{Yunes1}Yunes N and  Pretorius F 2009 Phys. Rev. D {\bf 79} 084043

\bibitem{Konno3} Konno K,  Matsuyama T and Tanda S 2009 Prog. Theor. Phys. {\bf122} 561




\bibitem{Cardoso} Cardoso V and Gualtieri L 2009 Phys. Rev. D {\bf 80} 064008

Molina C, Pani P,  Cardoso V and Gualtieri L 2010 arXiv:1004.4007




\bibitem{Ciufolini1} Ciufolini I 2007 arXiv: 0704.3338.
\bibitem{Ciufolini2} Ciufolini I and Pavlis E C 2004 Nature {\bf 431} 958



\bibitem{Yunes3} Yunes N and Spergel D N 2009 Phys. Rev. D {\bf 80} 042004

\bibitem{Harko} Harko T, Kov\'{a}cs Z and Lobo F S N 2010 Class. Quant. Grav. {\bf 27} 105010


\bibitem{Amarilla}Amarilla L, Eiroa E F and Giribet G 2010 Phys. Rev. D {\bf81} 124045
\bibitem{Ahmedov1}Ahmedov H and Aliev A N 2010 Phys. Lett. B {\bf690} 196

Ahmedov H and  Aliev A. N, 2010 Phys. Rev. D{\bf82} 024043


\bibitem{Carlos}Sopuerta  C F and Yunes N 2009 Phys. Rev. D {\bf80} 064006




\bibitem{Bozza2}Bozza V 2002 Phys. Rev. D {\bf 66} 103001

\bibitem{Bozza3} Bozza  V 2003 Phys. Rev. D {\bf 67} 103006

Bozza V, De Luca F, Scarpetta G and Sereno M 2005 Phys. Rev. D {\bf
72} 08300

Bozza V, De Luca F and Scarpetta G 2006 Phys. Rev. D {\bf 74} 063001


\bibitem{Harko2} Harko T, Kovacs Z and  Lobo F S N 2008 Phys. Rev. D {\bf78} 084005

Harko T, Kovacs Z and  Lobo F S N 2009 Phys. Rev. D {\bf79} 064001

\bibitem{Ksq} Matsuno K and Ishihara H 2009 Phys. Rev. D {\bf80} 104037





\bibitem{Vir2}Claudel C M, Virbhadra K S and Ellis  G F R 2001 J. Math. Phys. {\bf 42} 818

\bibitem{Ein1} Einstein A 1936 Science {\bf84} 506


\bibitem{Gyulchev} Gyulchev G N and Yazadjiev S S 2007 Phys. Rev. D {\bf75} 023006


\bibitem{Gyulchev1} Gyulchev G N and Yazadjiev S S 2008 Phys. Rev. D {\bf78}
083004

\bibitem{Darwin} Darwin C 1959 Proc. of the Royal Soc. of London {\bf 249} 180

\bibitem{Vir}Virbhadra K S,  Narasimha D and Chitre S M 1998 Astron. Astrophys. {\bf 337} 18

\bibitem{Vir1}Virbhadra K S and Ellis G F R 2000 Phys. Rev. D {\bf 62} 084003



\bibitem{Vir3} Virbhadra K S and Ellis G F R 2002 Phys. Rev.D {\bf 65} 103004


\bibitem{Fritt} Frittelly S, Kling T P and Newman E T 2000 Phys. Rev. D {\bf 61} 064021

\bibitem{Bozza1}Bozza V, Capozziello S, lovane G and
Scarpetta G 2001 Gen. Rel. and Grav. {\bf 33} 1535

\bibitem{Eirc1}Eiroa E F,  Romero G E and Torres D F 2002 Phys. Rev. D {\bf 66}
024010

Eiroa E F 2005 Phys. Rev. D {\bf 71} 083010

Eiroa E F 2006 Phys. Rev. D {\bf 73} 043002

\bibitem{whisk} Whisker R 2005 Phys. Rev. D {\bf71} 064004

\bibitem{Bhad1}Bhadra A 2003 Phys. Rev. D {\bf 67} 103009

\bibitem{Song1} Chen S and Jing J 2009 Phys. Rev. D {\bf 80} 024036


\bibitem{Song2} Liu Y, Chen S and Jing J 2010 Phys. Rev. D {\bf81} 124017

\bibitem{TSa1} Ghosh T andSengupta S 2010 Phys. Rev. D {\bf81} 044013

\bibitem{AnAv}Aliev A N and Talazan P 2009 Phys. Rev. D {\bf80} 044023


\bibitem{Genzel1}Genzel R, Eisenhauer F and Gillessen S 2010
arXiv:1006.0064

\bibitem{Burgay} Burgay M \textit{ et al} 2003 Nature. {\bf426} 531



\end{thebibliography}
\end{document}